\documentclass[fleqn,10pt]{wlscirep}

\usepackage[utf8]{inputenc}
\usepackage[T1]{fontenc}
\usepackage{graphicx} 
\usepackage{subcaption}
\usepackage{mathtools}
\usepackage{xcolor} 
\usepackage{comment} 
\usepackage{algorithm}
\usepackage[noend]{algpseudocode}
\usepackage{amsmath}
\usepackage{amsfonts} 
\usepackage{array}
\usepackage{graphicx} 

\title{ Publication Trends in Artificial Intelligence Conferences: The Rise of Super Prolific Authors}

\author[1,*]{Ariful Azad}
\author[1]{Afeefa Banu}

\affil[1]{Department of Intelligent Systems Engineering, Indiana University, Bloomington, IN USA}

\affil[*]{Corresponding author:azad@iu.edu}

\begin{abstract}

Papers published in top conferences contribute influential discoveries that are reshaping the landscape of modern Artificial Intelligence (AI). We analyzed 87,137 papers from 11 AI conferences to examine publication trends over the past decade. Our findings reveal a consistent increase in both the number of papers and authors, reflecting the growing interest in AI research. We also observed a rise in prolific researchers who publish dozens of papers at the same conference each year. In light of this analysis, the AI research community should consider revisiting authorship policies, addressing equity concerns, and evaluating the workload of junior researchers to foster a more sustainable and inclusive research environment.
\end{abstract}

\begin{document}
\flushbottom
\maketitle

\section{Introduction}
Artificial Intelligence (AI) has become one of the most intensively researched topics in recent years. 
Its rapid rise is driven by advancements in deep learning technology, which is projected to become a cornerstone of future economic growth, contributing tens of trillions of dollars globally in the near future~\cite{chui2023economic, russell2022human}.

AI's influence spans across diverse scientific fields, with a growing proportion of research focusing on applications of AI in areas such as healthcare, climate science, and more~\cite{rolnick2022tackling, ching2018opportunities, mater2019deep}.
However, central to AI's progress are its major conferences, where research undergoes rigorous peer review and is shared with the global community. Key venues include NeurIPS, AAAI, ICML, ICLR, IJCAI for general AI; CVPR, ICCV for computer vision; EMNLP, ACL for natural language processing; KDD for data mining; and ACM CHI for human-computer interaction.

From 2014 to 2023, these 11 premier conferences published an extraordinary 87,137 papers that significantly shaped AI's trajectory. Groundbreaking work from these venues includes advances in image classification~\cite{he2016deep, krizhevsky2012imagenet}, language models~\cite{brownlanguage}, and architectures like transformers and graph neural networks~\cite{vaswani2017attention, kipf2017semi, sutskever2014sequence}. These influential papers have garnered citations in the hundreds of thousands, underscoring their profound impact on the scientific community.

This paper analyzed 87,137 papers published across 11 top AI conferences over the last decade to determine publication trends. 
Our study investigates growth patterns in both the number of papers and participating authors, highlighting the emergence of prolific authors publishing dozens of papers in the same conferences. 
Through this analysis, we aim to uncover potential inequities in publishing opportunities, shedding light on how conference policies and practices may influence the distribution of research contributions among the global AI community.

Several studies have explored prolific authors in various disciplines, such as Accounting~\cite{hasselback2003prolific}, Medicine~\cite{wager2015too}, and broader academic topics~\cite{ioannidis2018thousands, ioannidis2024evolving}. 
However, the volume of papers published in AI far exceeds that of other fields, reflecting the rapid growth and interest in this area. This paper contributes to understanding publication trends in AI research, offering insights into the unique challenges and opportunities associated with the scale and dynamics of AI conferences.

\section{Publication trend in AI/ML conferences}
The growing interest and applications of Artificial Intelligence (AI) and Machine Learning (ML) are clearly reflected in the leading conferences in these fields. We identified 11 major conferences that have published the highest number of papers in recent years. These include conferences in general AI (NeurIPS, AAAI, ICML, ICLR, IJKAI), computer vision (CVPR, ICCV), natural language processing (EMNLP, ACL), data mining (KDD), and human-computer interaction (ACM CHI).
From 2014 to 2023, these 11 conferences collectively published an impressive total of 87,137 papers.
We collected information about these papers from conference websites and the DBLP website.

\begin{figure}[!t]
    \centering
    \includegraphics[width=\linewidth]{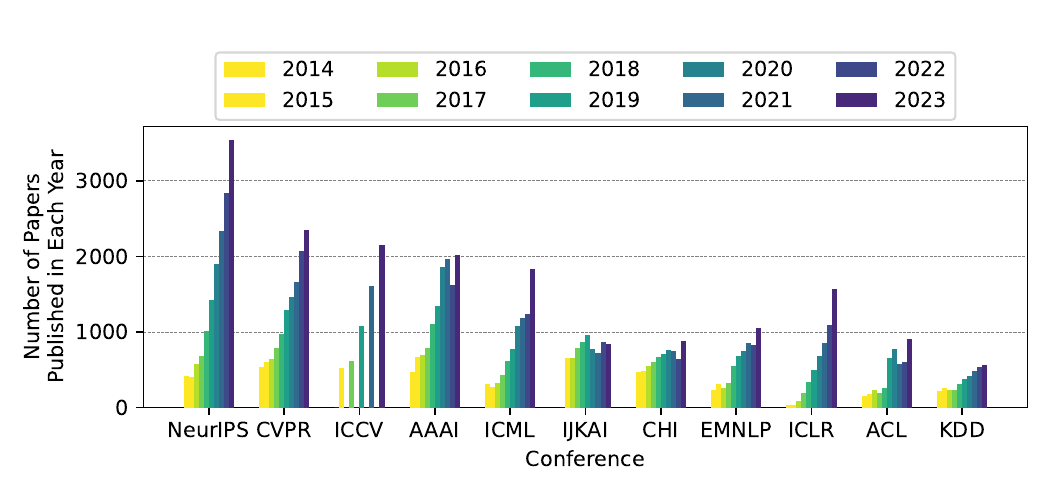}
    \vspace{-5pt}
    \caption{The number of papers published each year in top AI and ML conferences. }
    \label{fig:papers_per_year}
\end{figure}

\begin{figure}[!t]
    \centering
    \includegraphics[width=\linewidth]{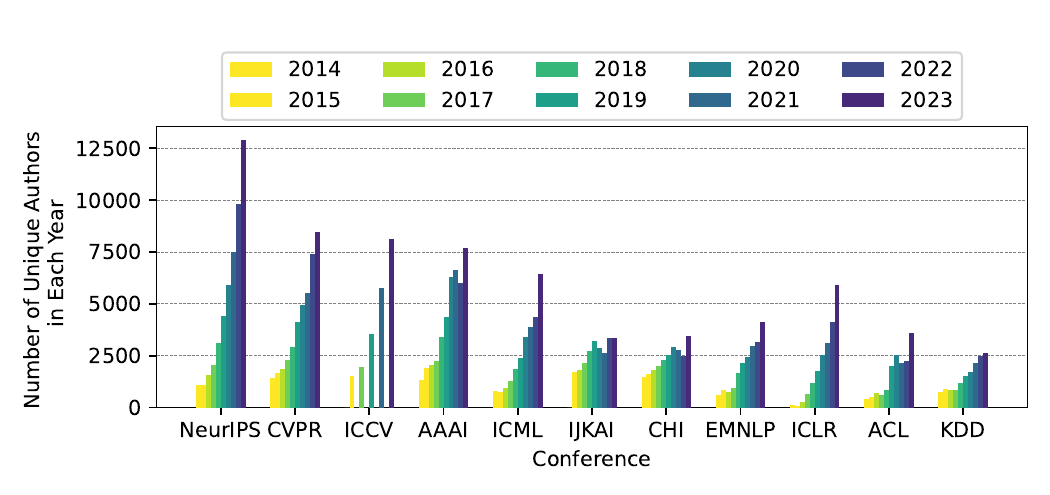}
    \vspace{-5pt}
    \caption{The number of unique authors publishing at least one paper each year in top AI and ML conferences. We counted distinct authors for each conference per year, ensuring that each author was only included once per conference annually.}
    \label{fig:authors_per_year}
\end{figure}

Across all AI conferences, a clear trend emerges: the number of accepted papers has consistently increased over time. 
Figure~\ref{fig:papers_per_year} illustrates the growth in paper acceptance for major AI conferences over the past decade.
While most conferences demonstrate a steady rise in the number of published papers, a notable exception is the International Joint Conferences on Artificial Intelligence (IJCAI).
Unlike others, IJCAI has stabilized its publication volume in recent years. 
We attribute this stability to strict policy, which limits authors to submitting a maximum of eight papers per year. 
This policy likely curtails excessive submissions and fosters a more balanced participation. 

Figure~\ref{fig:papers_per_year} highlights a similar trend in the number of authors publishing in major conferences over the past decade. For instance, in NeurIPS 2023, over 13,000 unique authors contributed at least one paper.
However, IJCAI stands out once again with a stabilized number of contributing authors, mirroring its trend of maintaining consistent paper acceptance rates.

\section{The rise of prolific authors in AI/ML conferences}
As in any social or economic development process, a fundamental question arises in AI research: does the growth of AI publications foster balanced participation among researchers? To explore this, we define an author as ``prolific" if they publish at least five papers in the same conference within the same year. Prolific authors are counted separately for each conference and year.
Although the threshold of five is arbitrary, achieving this milestone typically entails producing around 50 pages of robust scientific content, excluding appendices. This makes it a reasonable benchmark for defining prolific authorship.
Our choice of this threshold is inspired by IJCAI's submission policy, which limits authors to a maximum of eight submissions per year. This cap translated to at most five accepted papers per author, offering a balanced and fair model for prolific contributions.

\begin{figure}[!t]
    \centering
    \includegraphics[width=\linewidth]{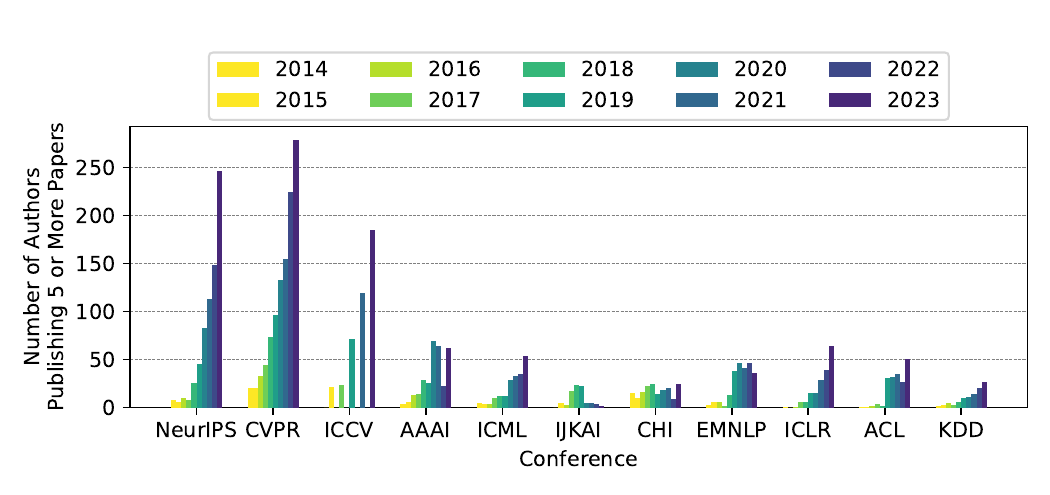}
    \caption{The number of authors publishing five or more papers annually in top AI and ML conferences. This reveals trends in prolific authorship across these venues.}
    \label{fig:prolific_authors_by_year}
\end{figure}

\begin{figure}[!t]
    \centering
    \includegraphics[width=\linewidth]{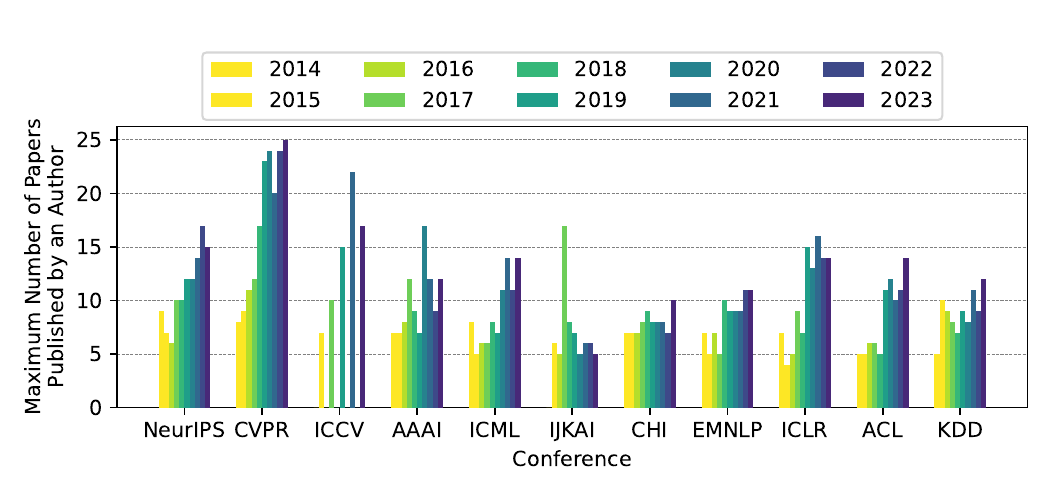}
    \caption{The number of papers published by the most prolific author in a given conference. The most prolific authors are counted separately for different conferences and years.}
    \label{fig:most_prolific_author_conf}
\end{figure}

Figure~\ref{fig:prolific_authors_by_year} illustrates the rising trend in the number of prolific authors across most AI conferences. For instance, NeurIPS 2023 and CVPR 2024 each saw over 250 authors publishing at least five papers.
This trend becomes even more striking when examining the maximum number of papers published by a single author in a conference, as shown in Figure~\ref{fig:most_prolific_author_conf}. In recent years, some authors have contributed over 20 papers to CVPR. We note that the most prolific author in a given year may differ from year to year.
Figure~\ref{fig:most_prolific_author_conf} further reveals that nearly all conferences, with the exception of IJKAI, allow individual authors to publish more than 10 papers annually within the same conference. This exception aligns with IJKAI's submission cap, which restricts the number of papers a single author can submit.

\begin{figure}[!t]
    \centering
    \includegraphics[width=.8\linewidth]{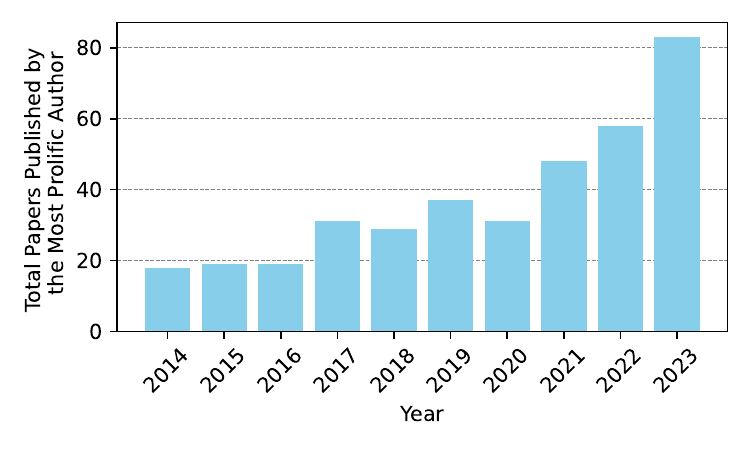}
    \caption{The total number of papers published by the most prolific author in a given calendar year across 11 AI conferences highlights their exceptional contribution to the field.}
    \label{fig:paper_by_most_prolific_author_year}
\end{figure}

Next, we identify the most prolific author in each calendar year, considering only papers published in the 11 conferences included in our analysis.
As shown in Figure~\ref{fig:paper_by_most_prolific_author_year}, the most prolific author each year has published an extraordinary number of papers in top AI conferences. For instance, one author contributed over 80 papers across the 11 conferences in 2023. Notably, this count excludes any papers these authors may have published in other venues.
The scale of this achievement is remarkable -- 80 papers equate to over 1,000 pages of high-quality scientific writing, including rigorous research and analysis. Such prolific output underscores the intensive effort and resources required to achieve these milestones in AI research.

\section{The equity in publications in AI conferences}
Given the dramatic rise of prolific authors in AI conferences, we analyzed the fraction of papers co-authored by these researchers. Specifically, we counted the number of papers authored by individuals who contributed to at least five papers within the same conference in a given year. 
We then calculated the fraction of these papers relative to the total number of papers published in the conference.
Additionally, we computed the fraction of prolific authors in each conference by dividing the number of authors publishing five or more papers by the total number of authors contributing to the conference.

\begin{figure}[!t]
    \centering
    \includegraphics[width=\linewidth]{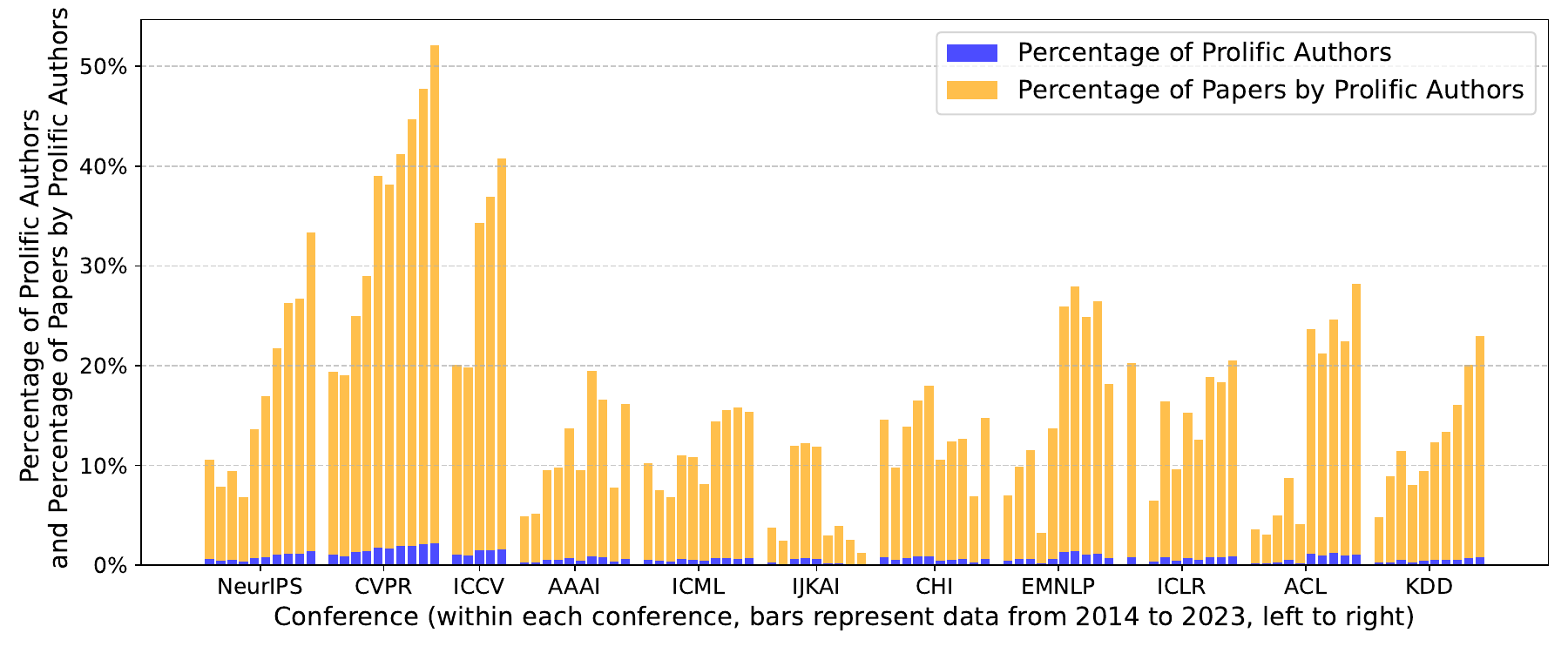}
    \caption{Fraction of prolific authors who published 5 or more papers in the same conference in a given year.}
    \label{fig:fraction_by_prolific_authors}
\end{figure}

Figure~\ref{fig:fraction_by_prolific_authors} reveals that prolific authors have contributed a significant share of the papers in recent conferences. A particularly striking example is CVPR 2023, where just 1\% of authors were contributed for over 50\% of the published papers.
In contrast, IJCAI exhibits more balanced contributions, with prolific authors contributing to less than 3\% of total papers in recent years. This difference highlights a notable disparity in author equity among leading conferences.
The increasing trend observed in conferences such as NeurIPS, CVPR, and ICCV suggests that this inequity may continue to increase in the coming years unless proactive measures are taken to address it.

\section{Discussion}
\subsection{Is the Publication Trend in AI Beneficial for the Scientific Community?}
In a recent article titled ``Surge in number of ‘extremely productive’ authors concerns scientists", Conroy expressed concerns about the rise of super-prolific authors across various scientific fields, including clinical medicine, agriculture, fisheries, and forestry~\cite{conroy2024surge, ioannidis2024evolving}. Similarly, Wager et al. expressed alarm upon identifying 24 authors listed in at least 25 publications in a single year within the field of Accounting~\cite{wager2015too}.
Given these concerns across multiple disciplines, it is timely to critically examine whether the current publication practices in AI conferences are healthy for the global AI community.

\subsection{A Focus on Authorship Policy}
Given the significant number of papers a single author can publish in a conference, questions arise about the adequacy of their contributions to each paper. While various publication venues adhere to different authorship policies, the Association for Computing Machinery (ACM) stipulates that every author must make significant intellectual contributions to some component of the original work described in the manuscript and take full responsibility for all content in the published work.
In light of this policy, it is debatable whether a researcher can make substantial contributions to 25 papers published in the same conference in a single year.

There is growing concern about improper authorship in scientific publishing, including the practice of selling authorships in exchange for money~\cite{else2023multimillion}. Given the prestige and potential financial benefits of being an author at top AI conferences, similar concerns apply to papers published in AI conferences.
To address some of these concerns, most AI conferences prohibit altering the author list or order after paper acceptance. However, a stricter policy may be necessary to ensure that authorship is reserved for those who have made significant contributions to the paper.

\subsection{A Focus on Workload of Graduate Students and Junior Researchers}
It is undeniable that much of the development and analysis in research is carried out by students and junior researchers. With the ever-increasing number of papers published by various research groups, junior researchers are often overwhelmed by continuous deadlines for multiple conferences. This raises concerns about their mental health and work-life balance.
Several studies have examined the well-being of engineering graduate students~\cite{bork2022engineering}, but the publication pressure appears to be particularly pronounced in AI research. 
Given the unique challenges faced by junior researchers in this field, it is essential to study the workload and stress factors associated with submitting papers to AI conferences. Addressing these concerns is crucial for promoting a healthier and more sustainable research environment for the next generation of researchers.

\subsection{A Focus on Equity and Inclusion}
A fundamental question in the social sciences is whether disparities in controlling a society's resources hinder economic growth~\cite{bagchi2015does}. For example, a recent study revealed that since 2020, the richest 1\% have captured nearly two-thirds of all new wealth—almost twice as much as the bottom 99\% of the global population~\cite{christensen2023survival}.
Our analysis shows a similar disparity in AI publications, where approximately 1\% of researchers contributed to more than 50\% of the papers in certain conferences. If this trend persists, it could have a negative impact on equity and inclusion in AI research.

\section{Analysis Methods}
We analyzed conference authorship data using Python libraries such as NetworkX, Matplotlib, and Pandas. The data and code are available upon request.

\section*{Author Contributions} 
AB collected data from conference websites and developed the initial analysis pipeline. AA completed the analysis and drafted the initial version of the paper. Both authors contributed to the finalization of the manuscript.

\section*{Competing interests}
The authors declare no competing interests.

\section*{Acknowledgements}
This work was supported by NSF grant OAC-2339607 and DOE grant DE-SC0022098.

\section*{Materials \& Correspondence} Correspondence should be sent to azad@iu.edu

\bibliography{bibfile}
\end{document}